\documentclass[conference]{IEEEtran}
\IEEEoverridecommandlockouts
\usepackage{setspace,cite}
\usepackage[overload]{empheq} %align in "cases"
\usepackage{graphicx}
\usepackage{epstopdf}
\usepackage{graphics} % for pdf, bitmapped graphics files
\usepackage{epsfig} % for postscript graphics files
\usepackage{color}
\usepackage{amsmath}%mathematic equations: aligned environment
\usepackage{amssymb} %\mathbb
\usepackage{amsthm}
\usepackage{verbatim}  %multiline comments
\usepackage{mathtools}%\DeclarePairedDelimiter
\usepackage{subfig}%subfigure
\usepackage{tabularx}
\usepackage{amsthm}
\usepackage{cases}%the use of numcases: numbering cases
\usepackage{verbatim}  %multiline comments
\usepackage{relsize}% resize sum sign
\newtheorem{theorem}{Theorem}%[section]
\newtheorem{lemma}{Lemma}%\newtheorem{lemma}[theorem]{Lemma}

\newtheorem{definition}{Definition}%[section]
\newtheorem{remark}{Remark}%[section]
\usepackage[ colorlinks = true,
linkcolor = blue,
urlcolor = blue,
citecolor = red,
anchorcolor = green,
]{hyperref}
\DeclarePairedDelimiter{\ceil}{\lceil}{\rceil}

\def\mcl{\mathcal}

\allowdisplaybreaks

\begin{document}
\title{Privacy Under Hard Distortion Constraints\\
%\title{Privacy Under Hard Distortion Constraints\textbf{}\\
}

\author{\IEEEauthorblockN{Jiachun Liao, Oliver Kosut, Lalitha Sankar}
	\IEEEauthorblockA{School of Electrical, Computer and Energy Engineering,\\
		Arizona State University\\
		Email: \{jiachun.liao,lalithasankar,okosut\}@asu.edu}
	\and
	\IEEEauthorblockN{Flavio P. Calmon}%Flavio du Pin Calmon
	\IEEEauthorblockA{School of Engineering and Applied Sciences\\
		Harvard University\\
		Email: fcalmon@g.harvard.edu\\
	}
	\thanks{This material is based upon work supported by the National Science Foundation under Grant No. CCF\--1350914 and CIF-1422358.}
	%\thanks{This material is based upon work supported by the National Science Foundation under Grant No. CCF\--1350914 and CIF\--1422358.}
}

%\and
%\and
%\and
%\and
%\and

\maketitle

\begin{abstract}
We study the problem of data disclosure with privacy guarantees, wherein the utility of the disclosed data is ensured via a \emph{hard distortion} constraint. Unlike average distortion, hard distortion provides a deterministic guarantee of fidelity. For the privacy measure, we use a tunable information leakage measure, namely \textit{maximal $\alpha$-leakage} ($\alpha\in[1,\infty]$), and formulate the privacy-utility tradeoff problem. The resulting solution highlights that under a hard distortion constraint, the nature of the solution remains unchanged for both local and non-local privacy requirements. More precisely, we show that both the optimal mechanism and the optimal tradeoff are invariant for any $\alpha>1$; i.e., the tunable leakage measure only behaves as either of the two extrema, i.e., mutual information for $\alpha=1$ and maximal leakage for $\alpha=\infty$.	
	
%In the study of a data disclosure problem under privacy protection, we enforce utility via a \emph{hard distortion} constraint, wherein the distortion between the original and disclosed data sets is upper bounded with probability $1$. Unlike average distortion, hard distortion provides a deterministic guarantee of fidelity. We use a tunable information leakage measure, namely \textit{maximal $\alpha$-leakage} ($\alpha\in[1,\infty]$), as the privacy metric and hard distortion as the utility measure to formulate a privacy-utility tradeoff (PUT) problem. The resulting solution highlights that under a hard distortion constraint, the nature of the solution remains unchanged for both local and non-local privacy requirements. More precisely, we show that both the optimal mechanism and the optimal tradeoff are invariant for any $\alpha>1$; i.e., the tunable leakage measure only behaves as either of the two extrema, i.e., mutual information (MI) for $\alpha=1$ and maximal leakage (MaxL) for $\alpha=\infty$.

\end{abstract}

\begin{IEEEkeywords}
Privacy-utility tradeoff, maximal $\alpha$-leakage, hard distortion, $f$-divergence.
\end{IEEEkeywords}

\section{Introduction}

From social networks to medical databases, useful cloud-based services require some form of user data disclosure to a third party. Data disclosure, however, often incurs a privacy risk. In most non-trivial settings, there is a fundamental trade-off between privacy and utility: on the one hand, disclosing data  ``as is'' can lead to unwanted inferences of private information. On the other hand, perturbing or limiting the disclosed data can result in a  reduced quality of service. 

The exact nature of the privacy-utility tradeoff (PUT) will depend to varying degrees on the distribution of the underlying data, as well as the chosen metrics (e.g., differential privacy \cite{Dwork_DP_Survey}, mutual information (MI) \cite{PrivacyAgainstStatistic_Calmon12,sankar_utility-privacy_2013}, $f$-divergence-based leakage measures \cite{TVDprivacy_Rassouli&Gunduz18}, maximal leakage (MaxL) \cite{MaximalLeakage_Issa2016}). Furthermore, most information-theoretic PUTs capture utility as a statistical average of desired measures of fidelity \cite{LocalPrivacy_Duchi13,StaircaseMechanismDP_Geng_Kairouz15,HypothesisTest&MutualInf_Liao2017,RobustPrivacy_Wang18}. This, in turn, simplifies the PUT to a single-letter optimization for independent and identically distributed (i.i.d.) datasets \cite{Privacy&MMSE_Asoodeh2016}. 
  
We measure utility in terms of a new \emph{hard distortion} metric, which constrains the privacy mechanism so that the distortion function between original and released datasets is bounded with probability $1$. This distortion metric is quite stringent, particularly when compared to average-case distortion constraints \cite{Privacy&MMSE_Asoodeh2016}, but it has the advantage that it allows the data curator to make specific, deterministic guarantees on the fidelity of the disclosed dataset to the original one. This differs significantly from a probabilistic constraint, which does not allow the data curator to make \emph{any} guarantee that the realization of the disclosed dataset has any relationship to the original one.

We adopt \textit{maximal $\alpha$-leakage}, which we introduced in \cite{ExtendedISIT18_Liao}, as an information leakage measure. Maximal $\alpha$-leakage is a \emph{tunable} privacy metric defined via an \textit{$\alpha$-loss} function with parameter $\alpha\in[1,\infty]$. For $\alpha=1$, this metric captures the inference gain by a (soft decision) belief-refining adversary after observing the disclosed data.  As $\alpha\to \infty$, this metric captures the reduction in $0-1$ loss or, equivalently, the gain of a (hard decision) adversary's guessing ability after data disclosure. These extreme points correspond to MI and MaxL, respectively. The tunable parameter $\alpha$ allows continuous interpolation between the two extremal adversarial actions by determining how much weight an adversary gives to its posterior belief.
 
Using the aforementioned utility and privacy measures, we precisely quantify the PUT and show that: (i) the same privacy mechanism achieves the same optimal PUT for all $\alpha>1$, and both the optimal mechanism and the optimal PUT are independent of the distribution of original data; (ii) For $\alpha=1$, the optimal privacy mechanism depends on the distribution of original data. More generally, for the sake of completeness, we also consider a larger class of $f$-divergence-based information leakages and derive the optimal PUTs for this class.

The paper is organized as follows: in Sec. \ref{Sec:privacy_metric}, we review maximal $\alpha$-leakage. In Sec. \ref{Sec:PUT_MaxAlphaLK-HardDist}, we formulate and solve the PUT problems with maximal $\alpha$-leakage as well as its $f$-divergence-based variants as privacy measures, and using hard distortion as the utility measure. In Sec. \ref{Section:Exp-PUT_HardDist_Types}, we illustrate our results via an example with binary data wherein the distortion function is the distance between types (empirical distributions) of the original and disclosed datasets.

\section{Maximal $\alpha$-Leakage and Related Leakage Measures}\label{Sec:privacy_metric}
Let $X$ and $Y$ represent the original and disclosed data, respectively, and let $U$ represent an arbitrary (potentially random) function of $X$ that the adversary (a curious or malicious observer of the disclosed data $Y$) is interested in learning. 
Maximal $\alpha$-leakage, introduced in \cite{ExtendedISIT18_Liao}, measures various aspects of leakage (ranging from the probability of correctly guessing to the posteriori distribution) about data $U$ from the disclosed $Y$. We review the formal definition next.

\begin{definition}[\hspace{-0.1pt}{\cite[Def. 5]{ExtendedISIT18_Liao}}]\label{Def:MaxAlphaLeakge}
	Given a joint distribution $P_{XY}$ on finite alphabets $\mcl X\times\mcl Y$, the maximal $\alpha$-leakage from $X$ to $Y$ is defined as
	\begin{align}
		&\mathsmaller{\mcl L_{\alpha}^{\text{max}}(X \to  Y)}\nonumber\\	
		\triangleq&\mathsmaller{\sup\limits_{U- X- Y}\lim\limits_{\alpha'\to \alpha}\frac{\alpha'}{\alpha'-1}\log\frac{\max\limits_{P_{\hat{U}|Y}}\mathbb{E}\left(\mathbb{P}(U=\hat{U}|U,Y)^{\frac{\alpha'-1}{\alpha'}}\right)}{\max\limits_{P_{\hat{U}}}\mathbb{E}\left(\mathbb{P}(U=\hat{U}|U)^{\frac{\alpha'-1}{\alpha'}}\right)}}, \label{eq:MaxAlphaLeak_definition}
	\end{align}
	where $\alpha\in[1,\infty]$, $U$ represents any function of $X$ and takes values from an arbitrary finite alphabet.
\end{definition}
Given an $\alpha$-loss function, $\alpha\in(1,\infty)$, 
\begin{equation}\label{poly_loss}
\ell_{\alpha}(u,y,P_{\hat{U}|Y})\triangleq\frac{\alpha}{\alpha-1} \big(1-P_{\hat{U}|Y}(u|y)^{1-\frac{1}{\alpha}}\big),
\end{equation} 
in the limits of $\alpha=1$ and $\alpha=\infty$ we obtain the log-loss and the 0-1 loss functions, respectively. 
A related $\alpha$-gain function of \eqref{poly_loss} is $1-\frac{\alpha-1}{\alpha}\ell_{\alpha}$. Therefore, maximal $\alpha$-leakage in \eqref{eq:MaxAlphaLeak_definition} measures the maximal multiplicative increase in the expected $\alpha$-gain for correctly inferring any function $U$ of $X$ when an adversary has access to $Y$ \cite{ExtendedISIT18_Liao}. The expression in \eqref{eq:MaxAlphaLeak_definition} can be further simplified to obtain the following theorem.

\begin{theorem}[\hspace{-0.1pt}{\cite[Thm. 2]{ExtendedISIT18_Liao}}]\label{Thm:MaxAlphaLeakge}
	For $\alpha\in[1,\infty]$, the maximal $\alpha$-leakage defined in \eqref{eq:MaxAlphaLeak_definition} simplifies to
\begin{align}
\mcl L_{\alpha}^{\text{max}}(X \to  Y)\hspace{2.2 in}\nonumber	
\end{align}
\begin{subnumcases}{=}
	 \sup\limits_{P_{\tilde{X}}}\inf\limits_{Q_Y}\,D_\alpha(P_{\tilde{X}}P_{Y|X}\|P_{\tilde{X}}\times Q_Y), &  $\alpha\in(1,\infty]$\hspace{20pt} \label{eq:GealLeak_EquivDef_alpha>1}\\
	 I(X;Y),   &   $\alpha=1$  \label{eq:GealLeak_EquivDef_alpha=1}
	\end{subnumcases}
%	\begin{align}
%	&\mcl L_{\alpha}^{\text{max}}(X \to  Y)\nonumber\\	
%     =&\begin{cases}
%	\sup\limits_{P_{\tilde{X}}}\inf\limits_{Q_Y}\,D_\alpha(P_{\tilde{X}}P_{Y|X}\|P_{\tilde{X}}\times Q_Y)\quad & \alpha\in(1,\infty]  \\
%	 I(X;Y)  &\alpha=1
%	\end{cases} \label{eq:GealLeak_EquivDef},
%	\end{align}
where in the supremum $P_{\tilde{X}}$ is constrained to have the same support as $X$, and $D_{\alpha}(\cdot)$ is the R{\'e}nyi divergence \cite{measures_renyi1961} of order $\alpha$ given by
\begin{align}
		D_{\alpha}(P_{\tilde{X}}P_{Y|X}\|P_{\tilde{X}}\times Q_Y)=\frac{1}{\alpha-1}
	\log\mathsmaller{\left(\sum\limits_{xy}\frac{P_{\tilde{X}}(x)P_{Y|X}(y|x)^{\alpha}}{Q_Y(y)^{\alpha-1}}\right)}\nonumber
\end{align}
and it is defined by its continuous extension for $\alpha=1$ or $\infty$.
\end{theorem}
The infimum over $Q_Y$ in \eqref{eq:GealLeak_EquivDef_alpha>1} is exactly Sibson MI of order $\alpha$\cite[Def. 4]{alphaMI_verdu}. Note that for $\alpha=1$ and $\alpha=\infty$, the maximal $\alpha$-leakage simplifies to MI and MaxL, respectively.
In \cite{ExtendedISIT18_Liao}, we show that maximal $\alpha$-leakage ($\alpha\in [1,\infty]$) satisfies data processing inequalities and a composition theorem.

While we are mainly interested in maximal $\alpha$-leakage, our results apply to a broader class of information leakages derived from $f$-divergences. 
Recall that for a convex function $f:\mathbb{R}\to \mathbb{R}$ such that $f(1)=0$, an $f$-divergence $D_f$ is a measure of the similarity between two distributions 
given by
\begin{align}\label{eq:f-divergence}
D_f(P\|Q) = \int dQ\, f\left(\frac{dP}{dQ}\right).
\end{align}
\begin{definition}\label{Def:Def-IntendedMeasure}
	Given a joint distribution $P_{XY}=P_{Y|X}P_X$ and a $f$-divergence $D_f$, a distribution-dependent leakage is defined as %on finite alphabets $\mcl X \times \mcl Y$ 
	\begin{align}\label{eq:Def-fLeakKforDist}
	\mcl L_f(X;Y)=\inf_{Q_Y} D_f(P_{XY}\|P_X\times Q_Y),
	\end{align} %footnotes: The independence is respect to the distribution of $X$. Different from that $\mcl L_f$ in \eqref{eq:Def-fLeakKforDist} depends on the distribution of $X$, the leakage $\mcl L_f^{\text{max}}$ in \eqref{eq:Definition-f_divergence_measure} only depends on the support of $X$.
	and a distribution-independent\footnote{The independence is with respect to the distribution of $X$. This ``distribution-independent'' measure depends on the distribution of $X$ only through its support. In contrast, the distribution-dependent measure $\mathcal{L}_f$ depends fully on the distribution of $X$. Both measures depend on the chosen mechanism $P_{Y|X}$.} leakage is defined as
	\begin{align}\label{eq:Definition-f_divergence_measure}
     \mcl L_f^{\text{max}}(X\to Y)=\sup_{P_{\tilde{X}}}\,\inf_{Q_Y}\, D_f(P_{\tilde{X}}P_{Y|X}\|P_{\tilde{X}}\times Q_Y),
	\end{align}
     where $P_{\tilde{X}}$ is constrained to have the same support as $P_X$.
\end{definition}
Recall that for $\alpha=1$, the maximal $\alpha$-leakage is MI and is a special case of $\mcl L_f(X;Y)$ in \eqref{eq:Def-fLeakKforDist} with $f(t)=t\log t$. Furthermore, for $\alpha>1$, maximal $\alpha$-leakage has a one-to-one relationship with a special case of $\mcl L_f^{\text{max}}$ in \eqref{eq:Definition-f_divergence_measure} for $f$ given by 
\begin{equation}\label{eq:Hellinger_f-function}
	f_\alpha(t)=\frac{1}{\alpha-1} (t^\alpha-1),
\end{equation}
such that $D_f$ is the Hellinger divergence of order $\alpha$ \cite{Liese2006}. The following lemma makes precise this observation.
\begin{lemma}\label{Lem:alphaLeakage_fDivergenceLeakage}
	For $\alpha>1$, maximal $\alpha$-leakage can be written as	
		\begin{align}	\label{eq:alphaLeakage_HellingerDivergenceLeakage}	
	\mcl L_{\alpha}^{\text{max}}(X\to Y)=\mathsmaller{\frac{1}{\alpha-1} \log \big(1+(\alpha-1) \mathcal{L}_{f_\alpha}^{\text{max}}(X\to Y)\big)},
		\end{align}
	where $\mathcal{L}_{f_\alpha}^{\text{max}}(X\to Y)$ is the $\mcl L_f^{\text{max}}(X\to Y)$ in \eqref{eq:Definition-f_divergence_measure}
	for $f_{\alpha}$ given by \eqref{eq:Hellinger_f-function} such that $D_f$ is the Hellinger divergence of order $\alpha$.
\end{lemma}

\section{Privacy-Utility Tradeoff with a Hard Distortion Constraint}\label{Sec:PUT_MaxAlphaLK-HardDist}
We now consider PUT problems minimizing either maximal $\alpha$-leakage or its related $f$-divergence-based variants in Def. \ref{Def:Def-IntendedMeasure}, subject to a \emph{hard} distortion constraint. Such a constraint can be written as $d(X,Y)\le D$ with probability $1$, where $d(\cdot,\cdot)$ is a distortion function and $D$ is the maximal permitted distortion. In other words, for any input $x\in \mcl X$, the output $y$ of the privacy mechanism must lie in a ball $B_D(x)$ given by
\begin{equation}
\label{eq:PUT_HardDist_CollectofFeasibleY}
B_D(x)\triangleq\{y:d(x,y)\le D\}.
\end{equation}

We henceforth denote an optimal PUT as $\text{PUT}_{\text{HD},\mcl L^{*}_{*}}$, where HD and $\mcl L^{*}_{*}$ in the subscript indicate the utility and privacy measures, respectively. 
The following two theorems characterize $\text{PUT}_{\text{HD},\mcl L_f}$ and $\text{PUT}_{\text{HD},\mcl L_f^{\text{max}}}$ with detailed proofs in appendices \ref{Proof:thm:PUT_fLeakKforDistvsHardDist} and \ref{Proof:Thm:PUT_fLeak_HardDist}, respectively. 
\begin{theorem}\label{thm:PUT_fLeakKforDistvsHardDist}
	For any distribution-dependent leakage $\mcl L_f$ in \eqref{eq:Def-fLeakKforDist} and a distortion function $d(\cdot,\cdot)$ with $B_D(x)$ in \eqref{eq:PUT_HardDist_CollectofFeasibleY}, the optimal PUT is given by
	{\small
		\begin{align}
		&\text{PUT}_{\text{HD},\mcl L_f} (D)\nonumber\\
		=&\inf_{P_{Y|X}: d(X,Y)\le D}\,\mcl L_f(X;Y) \label{eq:PUT_fLeakKforDistvsHardDist}\\		
		=&f(0)+\inf_{Q_Y} \mathbb{E} \mathsmaller{\left(Q_{Y}(B_D(X))\Big(f\big(\frac{1}{Q_Y(B_D(X))}\big)-f(0)\Big)\right)}.
		\label{eq:PUT-fLeakKforDist-HDdist-OptValue}
		\end{align}}
	If there exists a distribution $Q_Y^\star$ achieving the infimum in \eqref{eq:PUT-fLeakKforDist-HDdist-OptValue}, an optimal mechanism $P^*_{Y|X}$ is given by %\eqref{eq:opt_mech}.
	\begin{equation}\label{eq:opt_mech}
	\frac{dP_{Y|X=x}^*}{dQ_Y^\star}(y)=\frac{\mathbf{1}\big(d(x,y)\le D\big)}{Q_Y^\star(B_D(x))}.
	\end{equation}
	%\label{eq:MIvsHardDist_OptMech}
\end{theorem}
%As a result of the distribution dependence of the leakage measure $\mcl L_f$ in \eqref{eq:Def-fLeakKforDist}, the optimal tradeoff in \eqref{eq:PUT-fLeakKforDist-HDdist-OptValue} is an \textit{expectation} over the distribution of $X$.
\begin{figure}[t]
	\centering
	\includegraphics[width=3.5 in]{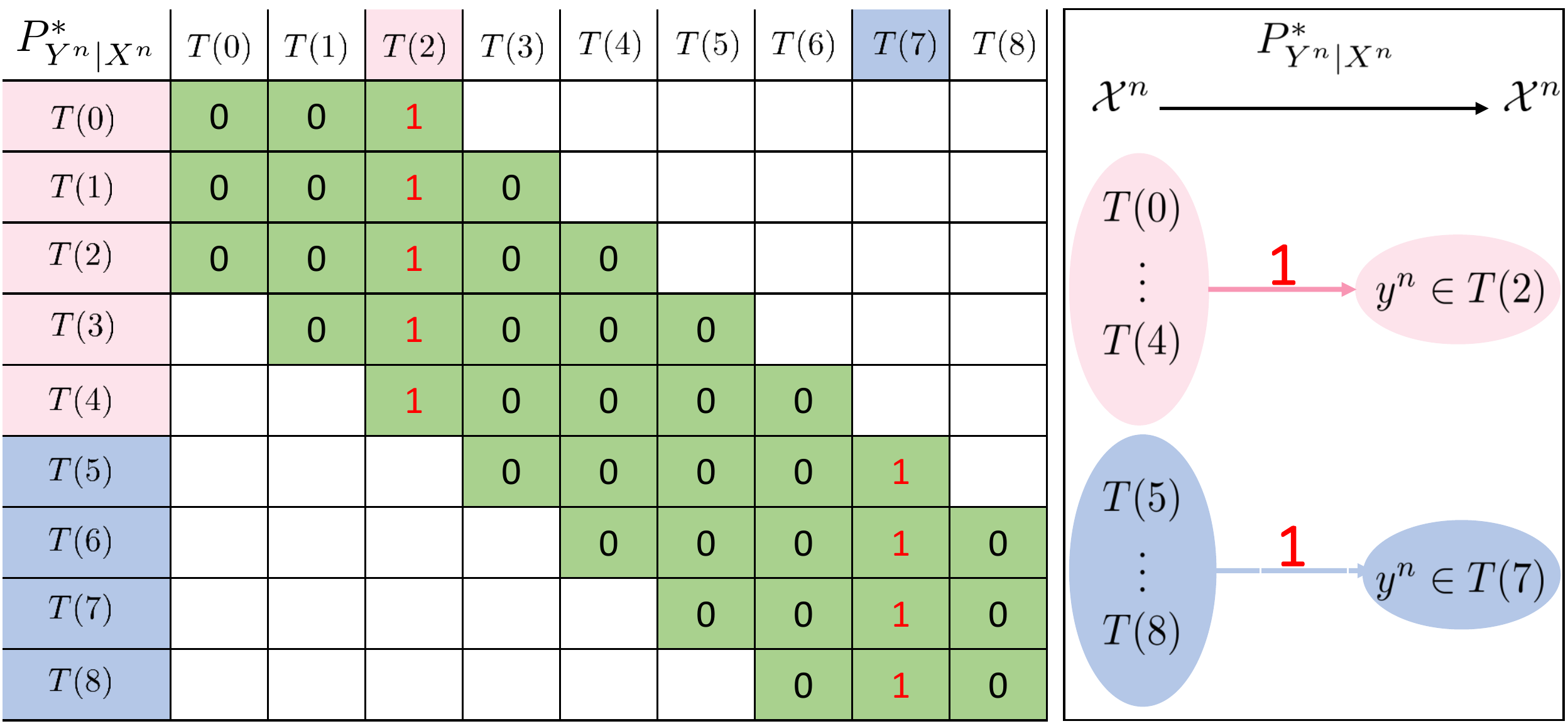}%Example_HardDist_XYChannelxy_1.pdf
	\caption{An optimal mechanism for $\alpha>1$ with $(n,m)=(8,2)$. Note that the hard distortion forces conditional probabilities of outputs outside the feasible ball of a given input to be zero. We highlight the conditional probabilities of feasible outputs in green, and give their values in the optimal mechanism.}
	%		\caption{An optimal mechanism of \eqref{eq:PUT_MaxAlphaLk&HardDist_Exp1} for $\alpha>1$ with $(n,m)=(8,2)$.}
	\label{fig:Example1_Channel_1}
	\vspace{-15 pt}
\end{figure}
\begin{theorem}\label{Thm:PUT_fLeak_HardDist}
	For any distribution-independent leakage $\mcl L_f^{\text{max}}$ in \eqref{eq:Definition-f_divergence_measure}, a distortion function $d(\cdot,\cdot)$ and $B_D(x)$ in \eqref{eq:PUT_HardDist_CollectofFeasibleY}, the optimal PUT is given by
	\begin{align}
	\text{PUT}_{\text{HD},\mcl L_f^{\text{max}}} (D)=&\inf_{P_{Y|X}:d(X,Y)\le D}\,\mcl L_f^{\text{max}}(X\to Y) 	\label{eq:PUT_fDivergenceLeak_HardDistortion}\\
	=&q^\star f((q^\star)^{-1})+(1-q^\star) f(0) \label{eq:PUT-fLeakK-HDdist-OptValue},
	\end{align}
	with $q^*$ defined as
	\begin{equation}\label{eq:q_star_def}
	q^\star\triangleq \sup_{Q_Y}\, \inf_x\, Q_Y(B_D(x)).
	\end{equation}
%    where $B_D(x)$ is defined in \eqref{eq:PUT_HardDist_CollectofFeasibleY}.
	Moreover, if there exists $Q_Y^\star$ achieving the supremum in \eqref{eq:q_star_def}, an optimal mechanism $P_{Y|X}^*$ is given by \eqref{eq:opt_mech}.
\end{theorem}
The PUTs in \eqref{eq:PUT-fLeakKforDist-HDdist-OptValue} and \eqref{eq:PUT-fLeakK-HDdist-OptValue} simplify to finding an output distribution $Q_Y$ that can be viewed as a ``target'' distribution, i.e., the optimal mechanism aims to produce this distribution as closely as possible, subject to the utility constraint. In particular, the resulting optimal mechanism (derived from \eqref{eq:opt_mech}), for any input, distributes the outputs according to $Q_Y$ while conditioning the output to be within a ball of radius $D$ about the input. The optimization in \eqref{eq:q_star_def} ensures that all inputs are uniformly masked while \eqref{eq:PUT-fLeakKforDist-HDdist-OptValue} provides average guarantees.

The following theorem characterizes the optimal tradeoff $\text{PUT}_{\text{HD},\mcl L_\alpha^{\text{max}}}$ for maximal $\alpha$-leakage. Recall that for $\alpha=1$, $\mcl L_{1}^{\text{max}}$ equals $\mcl L_f$ with $f(t)=t\log t$. For $\alpha>1$, from the one-to-one relationship between $\mcl L_\alpha^{\text{max}}$ and $\mcl L_{f_\alpha}^{\text{max}}$ in \eqref{eq:alphaLeakage_HellingerDivergenceLeakage}, we know that finding $\text{PUT}_{\text{HD},\mcl L_\alpha^{\text{max}}}$ is equivalent to finding the optimal tradeoff $\text{PUT}_{\text{HD},\mcl L_f^{\text{max}}}$ in \eqref{eq:PUT_fDivergenceLeak_HardDistortion} for $\mcl L_f^{\text{max}}=\mcl L_{f_\alpha}^{\text{max}}$. Due to space constraints, we omit details. %in Thm.~\ref{Thm:PUT_fLeak_HardDist} \eqref{eq:PUT_fDivergenceLeak_HardDistortion}
\begin{theorem}\label{Thm:PUT_MaximalAlphaLeak_HardDist}
	For maximal $\alpha$-leakage $\mcl L_\alpha^{\text{max}}$, a distortion function $d(\cdot,\cdot)$ and $B_D(x)$ in \eqref{eq:PUT_HardDist_CollectofFeasibleY}, the optimal PUT is given by 
	\hspace{-50pt}\begin{align}\label{eq:PUT_MaxAlphaLeak_HardDistortion}
	\text{PUT}_{\text{HD},\mcl L_\alpha^{\text{max}}} (D)=\inf_{P_{Y|X}: d(X,Y)\le D}\, \mcl L^{\text{max}}_\alpha(X\to Y)\,\qquad
	\end{align}
	\begin{subnumcases}{\hspace{+45pt}=}%\text{PUT}_{\text{HD},\mcl L_\alpha^{\text{max}}} (D)\hspace{-3pt}
	\inf_{Q_Y} \mathbb{E}\left(\log \mathsmaller{\frac{1}{Q_Y(B_D(X))}}\right),  \hspace{-15pt} & $\alpha=1$ \label{eq:PUT-MI-HDdist-OptValue}
	\\
	-\log q^\star,  \hspace{-15pt} & $\alpha>1$ \label{eq:PUT-MaxL-HDdist-OptValue}
	\end{subnumcases}
	where $q^\star$ is defined in \eqref{eq:q_star_def}. Moreover, an optimal mechanism is given by \eqref{eq:opt_mech}, where for $\alpha=1$, $Q_Y^\star$ achieves the infimum in \eqref{eq:PUT-MI-HDdist-OptValue}; and for $\alpha>1$, $Q_Y^\star$ achieves the supremum in \eqref{eq:q_star_def}.
\end{theorem}

\begin{remark}
		Note that subject to a hard distortion constraint, the optimal privacy mechanism is always given by \eqref{eq:opt_mech}. In particular, for maximal $\alpha$-leakage, the optimal mechanism as well as the optimal PUT are identical for all $\alpha>1$.
\end{remark}

\section{Example: Hard Distortion for Binary Types}\label{Section:Exp-PUT_HardDist_Types}
When considering dataset disclosure under privacy constraints, a reasonable goal is to design privacy mechanisms that preserve the statistics of the original dataset while preventing inference of each individual record (e.g., a sample or a row of the dataset). Since the type (empirical distribution) of a dataset captures its statistics, we quantify distortion as the distance between the type of the original and disclosed datasets. We use maximal $\alpha$-leakage to capture the gain of an adversary (with access to the disclosed dataset) in inferring any function of the original dataset.

Let $X^n$ be a random dataset with $n$ entries and $Y^n$ be the corresponding disclosed dataset generated by a privacy mechanism $P_{ Y^n|X^n}$. Entries of both $X^n$ and $Y^n$ are from the same alphabet $\mcl X$. For a pair of input and output datasets $(x^n,y^n)$ of $P_{Y^n|X^n}$, let $P_{x^n}$ and $P_{y^n}$ indicate the types, respectively. We define the distortion function as
\begin{equation}\label{eq:HDonTypes}
d(x^n,y^n)=\max_{x\in\mathcal{X}}|P_{x^n}(x)-P_{y^n}(x)|,
\end{equation}
and therefore, obtain $\text{PUT}_{\text{HD},\mcl L_\alpha^{\text{max}}} $ as in \eqref{eq:PUT_MaxAlphaLeak_HardDistortion} but with datasets $X^n,Y^n$ in place of single letters $X,Y$. Let the fraction $\frac{m}{n}$ ($m\in[0,n]$) be the upper bound $D$ in \eqref{eq:PUT_MaxAlphaLeak_HardDistortion}, where $[0,n]$ indicates the set of integers from $0$ to $n$.  

%For the simplification of expression, we describe the problem with type classes instead of datasets.
We concentrate on binary datasets and let $\mcl X=\{0,1\}$.
Note that for binary datasets, we can simply write $d(x^n,y^n)=|P_{x^n}(0)-P_{y^n}(0)|$.
For a $n$-length binary dataset, the number of types is $n+1$. Therefore, all input and output datasets can be categorized into $n+1$ type classes defined as  
   \begin{align}\label{eq:CollectionSeq_Type}
	T(i)\triangleq\{x^n:n P_{x^n}(0)=i\}=\{y^n:n P_{y^n}(0)=i\},\, i\in[0,n].\nonumber
	\end{align}
	
\begin{theorem}\label{Thm:PUTMaxAlphaLk&HardDist-SeqType-Alpha>1}
		Given an arbitrary pair of $(n,m)\in [1,\infty)\times [0,n]$, the minimal leakage for $\alpha>1$ is
		\begin{align}
		\text{PUT}_{\text{HD},\mcl L_\alpha^{\text{max}}} \left(\frac{m}{n}\right)=\log \mathsmaller{\ceil*{\frac{n+1}{2m+1}}}.
		\end{align}
		An optimal privacy mechanism maps all input datasets in \textit{a type class} to a unique output \textit{dataset} which is feasible and belongs to a type class in the set $\mcl T^*$ given by
		\begin{align}\label{eq:Opt-OutTypeSet}
		\mathsmaller{\mcl T^*\triangleq\Big\{T(j): j\hspace{-2pt}=\hspace{-2pt}l\hspace{-2pt}+\hspace{-2pt}(2m\hspace{-2pt}+\hspace{-2pt}1)k, k\in\mathsmaller{\left[0,\ceil*{\frac{n+1}{2m+1}}\hspace{-1pt}-\hspace{-1pt}1\right]}\Big\}},
		\end{align}
		where $l=m$ if $m+\left(\ceil{\frac{n+1}{2m+1}}-1\right)(2m+1)\leq n$, and otherwise, $l=n-\left(\ceil{\frac{n+1}{2m+1}}-1\right)(2m+1)$.
\end{theorem}
	A detailed proof is in Appendix \ref{Proof:Thm:PUTMaxAlphaLk&HardDist-SeqType-Alpha>1}. Let $(n,m)=(8,2)$ such that from Thm. \ref{Thm:PUTMaxAlphaLk&HardDist-SeqType-Alpha>1}, we have $\mcl T^*=\{T(2),T(7)\}$. Fig. \ref{fig:Example1_Channel_1} shows the optimal mechanism, which maps all input datasets in $\{T(i):i\in[0,4]\}$ (resp. $\{T(i):i\in[5,8]\}$) to a \textit{unique output dataset} in $T(2)$ (resp. $T(7)$) with probability $1$.

\section{Conclusion}
%\vspace{-5 pt}
We have explored PUTs in the context of hard distortion utility constraints. This utility constraint has the advantage that it allows the data curator to make specific, deterministic guarantees on the quality of the published dataset. Focusing on maximal $\alpha$-leakage and its $f$-divergence-based variants, under a hard distortion constraint, we have shown that: (i) for all $\alpha>1$, we obtain the same optimal privacy mechanism and optimal PUT, which are independent of the distribution of the original data (or datasets); (ii) for $\alpha =1$,  the optimal mechanism differs and depends on the distribution of the original data (or data sets). In other words, for this distortion measure, the tunable privacy measure behaves as either MI or MaxL. Possible future directions include verifying whether the observed behavior holds for average distortion constraints and more complicated data models.

\vspace{-5 pt}
\appendix
\vspace{-5 pt}
\subsection{Proof of Theorem \ref{thm:PUT_fLeakKforDistvsHardDist}}\label{Proof:thm:PUT_fLeakKforDistvsHardDist}
\vspace{-5pt}
\begin{proof}[\nopunct]%{eq:PUT_fDivergenceLeak_HardDistortion}
	The feasible ball $B_D(x)$ around $x$ is defined in \eqref{eq:PUT_HardDist_CollectofFeasibleY}. For the distribution dependent PUT in \eqref{eq:PUT_fLeakKforDistvsHardDist}, we have
	\begin{IEEEeqnarray}{l l}
	     	&\text{PUT}_{\text{HD},\mcl L_f} (D)\nonumber\\
			\label{eq:PUT_fLeakKforDistvsHardDist1}
			=&\inf_{P_{Y|X}: d(X,Y)\le D}\,\, \inf_{Q_Y}D_f(P_{Y|X}P_X\|P_X\times Q_Y)\\
			\label{eq:PUT_fLeakKforDistvsHardDist2}
			=&\inf_{Q_Y} \,\,\inf_{P_{Y|X}: d(X,Y)\le D} \int dP_{X}D_f(P_{Y|X=x}\|Q_Y)\\
			\label{eq:PUT_fLeakKforDistvsHardDist3}
			=&\inf_{Q_Y} \int \mathsmaller{dP_X} \inf_{\substack{P_{Y|X=x}\\Y\in B_D(x)}}\int dQ_{Y}f \left(\frac{d P_{Y|X}(\cdot|x)}{d Q_Y}\right)\\
			=&\inf_{Q_Y} \int \mathsmaller{dP_X} \inf_{\substack{P_{Y|X=x}\\Y\in B_D(x)}} \left(\int\limits_{y\in B_D(x)^c} dQ_{Y}f \left(\frac{d P_{Y|X}(\cdot|x)}{d Q_Y}\right)\right.\nonumber\\
			&\left.+Q_{Y}(B_D(x))\int\limits_{y\in B_D(x)} \frac{dQ_{Y}}{Q_{Y}(B_D(x))}f \left(\frac{d P_{Y|X}(\cdot|x)}{d Q_Y}\right)\right)\nonumber\\
			\label{eq:PUT_fLeakKforDistvsHardDist4}%
			\geq&\inf_{Q_Y} \int dP_X \inf_{\substack{P_{Y|X=x}\\Y\in B_D(x)}} \mathsmaller{\Big(Q_Y\left(B_D(x)^c\right) f(0)}\nonumber\\ &\quad\mathsmaller{\left.+Q_{Y}(B_D(x))f\left(\frac{1}{Q_Y(B_D(x))}\right)\right)}\\
			\label{eq:PUT_fLeakKforDistvsHardDist5}
			=& f(0)+\inf\limits_{Q_Y} \int  dP_X \Big(Q_{Y}(B_D(x))\big(f\big(\frac{1}{Q_Y(B_D(x))}\big)-f(0)\big)\Big)\nonumber
   \end{IEEEeqnarray}
	where \begin{itemize}
		\item \eqref{eq:PUT_fLeakKforDistvsHardDist2} follows from the fact that $D_f(P_{Y|X}\|Q_Y|P_X)$ is convex in $(P_{Y|X},Q_Y)$ for fixed $P_X$,
%		\item \eqref{eq:PUT_fLeakKforDistvsHardDist3} is from the fact that the conditional probability $P_{Y|X=dx}$ is independent to each other for different $x$,
		\item \eqref{eq:PUT_fLeakKforDistvsHardDist4} is from the Jensen's inequality and the equality holds if and only if there is a mechanism $P_{Y|X}$ satisfying
		\begin{align}\label{eq:OptMechvsQ_Y-InProof}
		\frac{d P_{Y|X}(y|x)}{d Q_Y(y)}=\frac{\mathbf{1}(y\in B_D(x))}{Q_Y(B_D(x))}.
		\end{align}
	\end{itemize}	
\end{proof}

\subsection{Proof of Theorem \ref{Thm:PUT_fLeak_HardDist}}\label{Proof:Thm:PUT_fLeak_HardDist}
\begin{proof}[\nopunct]
The feasible ball $B_D(x)$ around $x$ is defined in \eqref{eq:PUT_HardDist_CollectofFeasibleY}. For the distribution independent PUT in \eqref{eq:PUT_fDivergenceLeak_HardDistortion}, we have
	\begin{IEEEeqnarray}{l l}
		&\text{PUT}_{\text{HD},\mcl L_f^{\text{max}}} (D)\nonumber\\
		=&\inf_{P_{Y|X}:d(X,Y)\le D}\,\sup_{P_{\tilde{X}}}\,\inf_{Q_Y}\, D_f(P_{\tilde{X}}P_{Y|X}\|P_{\tilde{X}}\times Q_Y)\label{eq:put1}\\
		=& \inf_{Q_Y}\,\sup_{P_{\tilde{X}}}\,\inf_{P_{Y|X}:d(X,Y)\le D}\,D_f(P_{\tilde{X}}P_{Y|X}\|P_{\tilde{X}}\times Q_Y)\label{eq:put2}\\
		=&\inf_{Q_Y}\,\sup_{P_{\tilde{X}}}\,\inf_{\substack{P_{Y|X}\\d(X,Y)\le D}}\,\int \Big(dP_{\tilde{X}}(x)D_f(P_{Y|X=x}\|Q_Y)\Big)\label{eq:put3}\\
		= &\inf_{Q_Y}\, \sup_{P_{\tilde{X}}}\,\int dP_{\tilde{X}}(x)\mathsmaller{\inf\limits_{\substack{P_{Y|X=x}\\Y\in B_D(x)}}} \int \mathsmaller{dQ_Y f\left(\frac{dP_{Y|X=x}}{dQ_Y}\right)}\label{eq:put5}\\
		=&\inf_{Q_Y}\,\sup_{P_{\tilde{X}}}\, \int dP_{\tilde{X}}(x)\, \mathsmaller{\inf\limits_{\substack{P_{Y|X=x}\\Y\in B_D(x)}}}\Bigg(\int_{B_D(x)} dQ_Y\nonumber\\
		&\quad  \mathsmaller{f\left(\frac{dP_{Y|X=x}}{dQ_Y}\right)+\int_{B_D(x)^c} dQ_Y f(0)\Bigg)}\label{eq:put6}\\
		=&\inf_{Q_Y}\,\sup_{P_{\tilde{X}}}\, \int dP_{\tilde{X}}(x)\, \mathsmaller{\inf\limits_{\substack{P_{Y|X=x}\\Y\in B_D(x)}}\,
		\Bigg( Q_Y(B_D(x))} \int_{B_D(x)} \nonumber\\
		&\mathsmaller{\frac{dQ_Y}{Q_{Y}(B_D(x))}f\left(\frac{dP_{Y|X=x}}{dQ_Y}\right)+Q_Y(B_D(x)^c)f(0)\Bigg)}\label{eq:put7}\\
		\geq&\inf_{Q_Y}\,\sup_{P_{\tilde{X}}}\, \int dP_{\tilde{X}}(x)\, \mathsmaller{\inf\limits_{\substack{P_{Y|X=x}\\Y\in B_D(x)}}
		\bigg( Q_Y(B_D(x))} \nonumber\\%f(Q_Y(B_D(x))^{-1})
		&\quad \mathsmaller{\cdot f\left(\frac{1}{Q_Y(B_D(x))}\right)+\big(1-Q_Y(B_D(x))\big)f(0)\bigg)}\label{eq:put8}\\
		=&\inf_{Q_Y}\,\sup_{P_{\tilde{X}}}\,  \int dP_{\tilde{X}}(x)\, g\big(Q_Y(B_D(x))\big)\label{eq:put9}\\
		=&\inf_{Q_Y}\,\sup_x\, g\big(Q_Y(B_D(x))\big)\label{eq:put9+}
   \end{IEEEeqnarray}	
where \begin{itemize}\vspace{-6pt}
		\item \eqref{eq:put2} and \eqref{eq:put5} follow from the fact that $D_f(P_{XY}\|P_X\times Q_Y)$ is linear in $P_X$ for fixed $(P_{Y|X},Q_Y)$ and convex in $(P_{Y|X},Q_Y)$ for fixed $P_X$,
		\item \eqref{eq:put8} follows from the convexity of $f$ and Jensen's inequality. The equality holds if and only if there exists a mechanism $P_{Y|X}$ satisfying \eqref{eq:OptMechvsQ_Y-InProof}.
		\item \eqref{eq:put9} results from $q\triangleq Q_Y(B_D(x))$ and 
		\begin{equation}\label{eq:put10}
		g(q)\triangleq qf(q^{-1})+(1-q)f(0).
		\end{equation}
	\end{itemize}
    Due to the convexity of $f$, we have $f(q^{-1})-f(0)\le f'(q^{-1})\left(q^{-1} -0\right)$,
%	\begin{equation}
%%	f(q^{-1})-f(0)\le f'(q^{-1})\left(q^{-1} -0\right),
%	\end{equation}
	from which, the derivative $g'(q)=f(q^{-1})-q^{-1} f'(q^{-1})-f(0)\leq 0$. Therefore, the function $g$ in \eqref{eq:put10} is non-increasing, such that \eqref{eq:put9+} is be simplified as $g(q^\star)$, where $q^*$ is given by 
	\begin{equation}\label{eq:q_star_def-inPf}
		q^\star\triangleq \sup_{Q_Y}\, \inf_x\, Q_Y(B_D(x)).
	\end{equation}
\end{proof}

\subsection{Proof of Theorem \ref{Thm:PUTMaxAlphaLk&HardDist-SeqType-Alpha>1}}\label{Proof:Thm:PUTMaxAlphaLk&HardDist-SeqType-Alpha>1}
 \vspace{-10pt}
\begin{proof}[\nopunct]
	Define the feasible ball around an input dataset $x^n$ as \vspace{-5pt}
	\begin{align}
		B_D(x^n)\triangleq \left\{y^n: |P_{x^n}(0)-P_{y^n}(0)|\leq \frac{m}{n}\right\}.
	\end{align}
	From Thm. \ref{Thm:PUT_MaximalAlphaLeak_HardDist}, to find an optimal mechanism $P_{Y^n|X^n}^*$, we need to find an output distribution $Q_{Y^n}^*$ which optimizes \eqref{eq:q_star_def} with $x^n$ and $y^n$ in place of $x,y$.\\
	Note that for the hard distortion $|P_{x^n}(0)-P_{y^n}(0)|\leq \frac{m}{n}$, all datasets in a type class share the same group of feasible output datasets, and this feasible group can be represented by the type classes. Therefore, for any $x^n\in T(i)$ ($i\in[0,n]$), we rewrite $B_D(x^n)$ as \vspace{-3pt}
	\begin{align}
	B_D(x^n)=B_D(T(i))\triangleq\left\{T(j): \left|i-j\right|\leq m, j\in [0,n]\right\}.\nonumber
	\end{align} 
	We define an distribution $Q_{T}$ of type classes for outputs as  \vspace{-5pt}
	\begin{align}\label{eq:PUT-TYPE-RelatDisonTypeandSeq-inProof}
			Q_{T}(j)\triangleq\sum\limits_{y^n\in T(j)}Q_{Y^n}(y^n), \text{ for } j\in[1,n],
	\end{align}\vspace{-5pt}
	such that \vspace{-3pt}
	\begin{equation}\label{eq:q_star_def_TypeExamp}
	q^*=\sup_{Q_{T}}\, \inf_{i\in[0,n]}\, Q_{T}(B_D(T(i))).
	\end{equation}
	The optimal distribution $Q_{T}$ is determined by both upper and lower bounding $q*$ in \eqref{eq:q_star_def_TypeExamp}. The upper bound is determined by restricting the optimization in \eqref{eq:q_star_def_TypeExamp} to a judicious choice of a small set of input types. The lower bound is a constructive scheme. Let $l=m+(2m+1)\left(\ceil{\frac{n+1}{2m+1}}-1\right)-n$. We define an index set $\mcl I_T\subset [0,n]$ for types as 
	\small{\begin{align}\label{eq:index_types}
		I_T\triangleq\begin{cases}
			\left\{m+(2m+1)k: k\in\left[0,\ceil*{\frac{n+1}{2m+1}}-1\right]\right\} & l\leq 0\\
			\left\{l+(2m+1)k: k\in\left[0,\ceil*{\frac{n+1}{2m+1}}-1\right]\right\} & l> 0\\
		\end{cases}.
	\end{align}}\normalsize
	From the expression of $\mcl I_T$ in \eqref{eq:index_types}, we observe that: (i) for $l\leq 0$ (resp. $l>0$), the first (resp. last) element is $m$ (resp. $n$); (ii) for $l\leq 0$ (resp. $l>0$), the last (resp. first) element is no less (resp. less) than $n-m$ (resp. $m+1$); (iii) for both cases, the difference between adjacent elements is $2m+1$. Therefore, it is not difficult to see that feasible balls of input type classes indexed by $I_T$ are a partition of the set of all type classes, i.e., 
	 \vspace{-5pt}\begin{subequations}\label{eq:PartitionSets_inPf}
		\begin{align}
		    &B_D(T(i_1))\cap B_D(T(i_2))=\emptyset\quad i_1,i_2\in \mcl I_T,\\
			&\left\{T(j):j\in[0,n]\right\}  = \bigcup_{i\in\mcl I_T} B_D(T(i)).\vspace{-5pt}
		\end{align}
	\end{subequations}
 \vspace{-5pt}Therefore, the problem in \eqref{eq:q_star_def_TypeExamp} is upper bounded by
\begin{IEEEeqnarray}{l l}
			q^*\leq &\sup_{Q_{T}}\, \inf_{i\in\mcl I_T }\, Q_{T}(B_D(T(i)))\\
			\leq&\sup_{Q_{T}}\, \frac{1}{|\mcl I_T|} \sum_{i\in\mcl I_T }\, Q_{T}(B_D(T(i)))\\
			=& \sup_{Q_{T}}\mathsmaller{\left(\ceil*{\frac{n+1}{2m+1}}\right)^{-1}} \sum_{j\in[1,n]}Q_{T}(T(j))\\
			=&\mathsmaller{\left(\ceil*{\frac{n+1}{2m+1}}\right)^{-1}}.
\end{IEEEeqnarray}
Construct an distribution $Q'_{T}$ as \vspace{-5pt}% such that for every $j\in I_T$ and an unique $y^n\in T(j)$
\begin{align}\label{eq:PUT-TYPE-OptDisOnType}
Q_{T}'(j)=
\mathsmaller{\left(\ceil*{\frac{n+1}{2m+1}}\right)^{-1}}\quad \text{ for }j\in I_T,
\end{align}
and otherwise, $Q_{T}'(j)=0$.
%\begin{align}\label{eq:PUT-TYPE-OptDisOnType}
%	Q_{T}'(j)=\begin{cases}
%	\left(\ceil*{\frac{n+1}{2m+1}}\right)^{-1}& j\in I_T\\
%	0& \textit{otherwise}.
%	\end{cases}
%\end{align}
By \eqref{eq:PartitionSets_inPf} for each $i\in[0,n]$, there is a \textit{unique} $k$ satisfying $|i-I_T(k)|\leq m$, where $I_T(k)$ is the $k^{\text{th}}$ element\footnote{From \eqref{eq:index_types}, $I_T(k)$ is either $m+(2m+1)k$ or $l+(2m+1)k$.} of $I_T$.  Therefore, we lower bound \eqref{eq:q_star_def_TypeExamp} by 
\begin{IEEEeqnarray}{r l}
	q^*\geq &\,\inf_i Q_{T}'(B_D(T(i)))\\
	      = &\,\inf_i Q_{T}'\Big(\bigcup_{\substack{|i-j|\leq m\\j\in \mcl I_T}}T(j)\Big)\\
	      = &\,\inf_k Q_{T}'(T(I_T(k)))= \mathsmaller{\left(\ceil*{\frac{n+1}{2m+1}}\right)^{-1}}.
\end{IEEEeqnarray}  
Thus, $q^*=\left(\ceil*{\frac{n+1}{2m+1}}\right)^{-1}$ and the $Q_{T}'$ in \eqref{eq:PUT-TYPE-OptDisOnType} is optimal.\\
From \eqref{eq:PUT-TYPE-RelatDisonTypeandSeq-inProof} and the $Q_{T}'$, we derive an optimal $Q^*_{Y^n}$, which assigns the same non-zero probability to only one dataset of each type classes indexed by $I_T$, i.e., $Q^*_{Y^n}(y^n)=q^*$ for one $y^n\in T(j)$ for each $j\in I_T$. Therefore, from \eqref{eq:opt_mech} we have the corresponding optimal privacy mechanism, which maps all input datasets in one input type class to one feasible output dataset with probability $1$. %where $y^n\in T(I_T(k))$ with $|i-I_T(k)|\leq m$.

\end{proof}

\bibliographystyle{IEEEtran}
\bibliography{JL_References}
\end{document}